\documentclass[onecolumn,superscriptaddress,showpacs,showkeys]{revtex4}
\usepackage{epsfig}
\usepackage{amssymb,latexsym,amsmath}
\newcommand{\eq}[1]{\begin{align} #1 \end{align}}

\begin{document}

\title{ Particle Number Fluctuations
in High Energy Nucleus-Nucleus Collisions from
Microscopic Transport Approaches}

\author{V.P. Konchakovski}
\affiliation{Bogolyubov Institute for Theoretical Physics, Kiev, Ukraine}
\affiliation{Shevchenko National University, Kiev, Ukraine}
\author{S. Haussler}
\affiliation{Frankfurt Institute for Advanced Studies, Frankfurt, Germany}
\author{ M.I. Gorenstein}
\affiliation{Bogolyubov Institute for Theoretical Physics, Kiev, Ukraine}
\affiliation{Frankfurt Institute for Advanced Studies, Frankfurt, Germany}
\author{E.L. Bratkovskaya}
\affiliation{Frankfurt Institute for Advanced Studies, Frankfurt, Germany}
\author{ M. Bleicher}
\affiliation{Institut
f\"ur Theoretische Physik, Johann Wolfgang Goethe Universit\"at,
 Frankfurt, Germany}
\author{H. St\"ocker}
\affiliation{Frankfurt Institute for Advanced Studies, Frankfurt, Germany}
\affiliation{Institut f\"ur Theoretische Physik,
Johann Wolfgang Goethe Universit\"at,
 Frankfurt, Germany}

\begin{abstract}
Event-by-event multiplicity fluctuations in nucleus-nucleus collisions
are studied within the HSD and UrQMD transport models.  The scaled
variances of negative, positive, and all charged hadrons in Pb+Pb at
158 AGeV are analyzed in comparison to the data from the NA49
Collaboration. We find a dominant role of the fluctuations in the
nucleon participant number  for the final hadron multiplicity
fluctuations. This fact can be used to check different scenarios of
nucleus-nucleus collisions by measuring the final multiplicity
fluctuations as a function of collision centrality. The analysis
reveals surprising effects in the recent NA49 data which indicate a
rather strong mixing of the projectile and target hadron production
sources even in peripheral collisions.

\end{abstract}


\maketitle

\section{Introduction}

The aim of the present paper is to study particle number fluctuations
in high energy nucleus-nucleus (A+A) collisions within the HSD
\cite{HSD} and UrQMD \cite{UrQMD} transport approaches.  The analysis
of fluctuations is an important tool to study a physical system created
in high energy nuclear collisions.  Recently, preliminary NA49 data on
particle number fluctuations in Pb+Pb collisions at 158~A~GeV   for
different centralities have been presented \cite{NA49} which are in
surprising disagreement with the results of both microscopic transport
models that have been shown to reproduce both the different particle
multiplicities and longitudinal differential rapidity distributions for
central collisions of Au+Au (or Pb+Pb) collisions from AGS to SPS
energies rather well \cite{Weber}.

The fluctuations in high energy particle and nuclear collisions (see,
e.g.,
Refs.~\cite{fluc1,fluc2,fluc3,fluc4,fluc5,fluc6,fluc7,fluc7a,fluc7b,fluc8} and
references therein) are studied on an event-by-event basis:  a given
observable is measured in each event and the fluctuations are evaluated
for a  specially selected set of these events.  The statistical model
has been successfully used to describe the data on hadron
multiplicities in relativistic A+A collisions (see, e.g.,
Ref.~\cite{stat1} and a recent review \cite{BMST}) as well as in
elementary particle collisions \cite{stat2}. This gives rise to the
question whether the fluctuations, in particular the multiplicity
fluctuations, do also follow the statistical hadron-resonance gas
results. The statistical fluctuations can be closely related to phase
transitions in QCD matter, with specific signatures for  1-st and 2-nd
order phase transitions as well as for the critical point
\cite{fluc4,fluc5}.

In addition to the statistical fluctuations, the complicated dynamics of
A+A collisions generates {\it dynamical} fluctuations.  The
fluctuations in the initial energy deposited inelastically in the
statistical system yield {\it dynamical} fluctuations  of all
macroscopic parameters, like the total entropy or strangeness content. The
observable consequences of the initial energy density fluctuations are
sensitive to the equation of state of the matter, and can therefore
be useful as signals
for phase transitions \cite{fluc8}.  Even when the data are obtained with
a centrality trigger, the number of nucleons participating in
inelastic collisions still fluctuates considerably. In the language of
statistical mechanics, these fluctuations in participant nucleon number
correspond to volume fluctuations. Secondary
particle multiplicities scale linearly with the volume, hence,
volume fluctuations translate directly to particle number fluctuations.

In the present paper we study the particle number fluctuations in Pb+Pb
collisions at 158~AGeV within both the HSD and UrQMD transport models.
We check the robustness of the two approaches and derive physical
consequences from the results of the HSD and UrQMD simulations. Then we
formulate a general picture of particle number fluctuations in
different scenarios for A+A collision processes.  The paper is
organized as follows. Section II presents the HSD and UrQMD results in
comparison with NA49 data. Section III studies the role of the
fluctuations of the number of participant nucleons for the fluctuations
of the final hadron multiplicities. HSD and UrQMD calculations are
employed to clear up this point on a microscopic level.  Section IV
discusses a recently proposed method \cite{MGMG}, which allows to test
experimentally different model scenarios of A+A collisions. A
comparison of the model results to recent NA49 data shows a necessity
of strong mixing of the projectile and target hadron production sources
not only for central but also for more peripheral collisions. This
strong mixing is underestimated in the hadron/string dynamical
approaches. Section V finally presents our summary and conclusions.

\section{HSD and UrQMD results in comparison to the NA49 Data}

In each A+A event only a fraction of all 2$A$ nucleons (the participant
nucleons) interact. We denote the number of participant
nucleons from the projectile and target nuclei as $N_P^{proj}$ and
$N_P^{targ}$, respectively.  Those nucleons which do not interact are
called spectator nucleons. Their numbers are related to the participant numbers
as
$N_S^{proj} = A - N_P^{proj}$ and
$N_S^{targ} = A - N_P^{targ}$.
The trivial geometrical fluctuations due to impact parameter
variations usually dominate in high energy A+A collisions and mask
the fluctuations of interest. One cannot fix the impact parameter
experimentally, but even for a fixed impact parameter the number of
participants must fluctuate from event to event.  Moreover, the numbers of
the projectile and the target participants differ in a given event.
This is caused by fluctuations in the initial states of
the colliding nuclei and the probabilistic character of the various
hadron-hadron collision
processes.

The NA49 Collaboration has tried to minimize the event by event
fluctuations of the number of nucleon participants in measuring the
multiplicity fluctuations.  Samples of collisions with a fixed number
of projectile spectators, $N_S^{proj} = const$, and thus a fixed number
of projectile participants, $N_P^{proj} = A- N_S^{proj}$, were
selected.  This selection is possible in fixed target experiments,
where $N_S^{proj}$ is measured by a Zero Degree Veto Calorimeter, which
covers the projectile fragmentation domain.

  From an output of the HSD and UrQMD minimum bias
 simulations we form the samples of Pb+Pb events with fixed
 values of $N_P^{proj}$. In Fig. 1 we present the
 HSD and UrQMD results and compare
them with the NA49 data for the scaled variances of negatively, positively, and
all charged particles in Pb+Pb collisions at 158 AGeV. The average values
(we will use the double brackets to denote the averaging in the model
simulations),
$$\langle\langle N_i \rangle\rangle, \ \ (i=+,-,ch)$$
and variances
$$Var(N_i)\equiv \langle\langle N_i^2 \rangle\rangle-
       \langle\langle N_i\rangle\rangle^2$$
are calculated for the samples of collision events with fixed values of the
projectile participants, $N_P^{proj}$, and scaled variances are by
definition,
$$\omega_i\equiv Var(N_i)/\langle\langle N_i\rangle\rangle~.$$
Note that $\omega=1$ for the Poisson multiplicity distribution,
$P(N)=\exp(-\overline{N})\overline{N}^N/N!$~.

\begin{figure}[h!]
 \epsfig{file=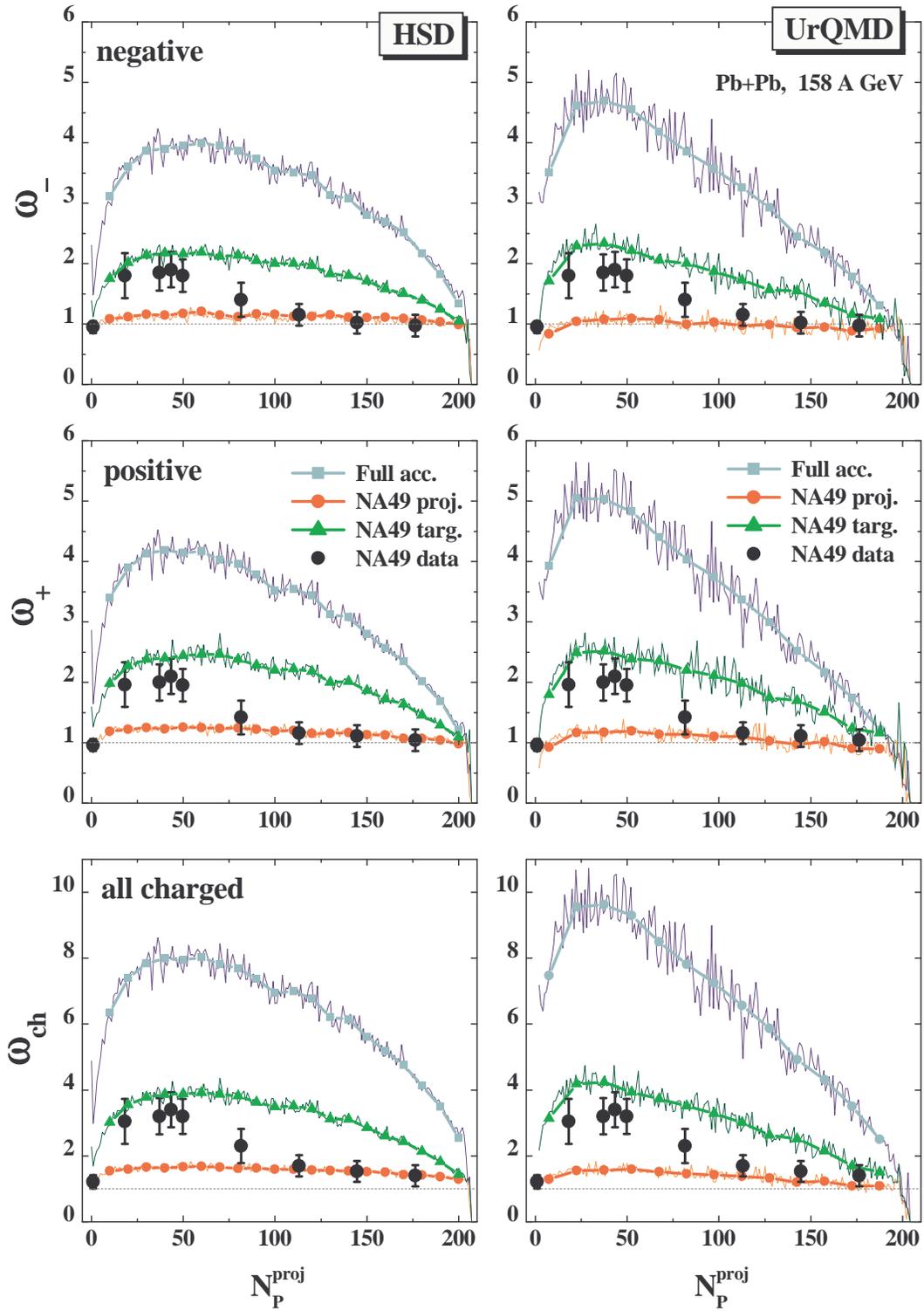,width=14cm}
  \caption{(Color online) The results of the HSD (left) and UrQMD (right) simulations
  are shown for $\omega_-$, $\omega_+$, and $\omega_{ch}$ in Pb+Pb collisions
  at 158~AGeV as functions of $N_P^{proj}$. The black points are the
  NA49 data. The different lines correspond to the model simulations
with the original NA49 acceptance, $1.1<y<2.6$, in the projectile hemisphere
  (lower lines), the NA49-like  acceptance in the
  mirror rapidity interval, $-2.6<y<-1.1$,
   in the target hemisphere (middle lines), and full $4\pi$ acceptance
   (upper lines).  }
 \label{w_Np}
 \end{figure}

The final particles in the HSD and UrQMD simulations are accepted
at rapidities $1.1< y <2.6$ (we use
particle rapidities in the Pb+Pb c.m.s. frame) in accord to the NA49
transverse momentum filter \cite{NA49}. This is done to compare
the HSD and UrQMD results with the NA49 data.
The HSD and UrQMD simulations both show flat $\omega_i$ values,
$\omega_-\approx \omega_+\approx 1.2$,
$\omega_{ch}\approx 1.5$, and exhibit almost no dependence on $N_P^{proj}$.
The NA49 data, in contrast, exhibit  an enhancement in $\omega_i$
 for $N_P^{proj}\approx
50$. The data show maximum values, $\omega_-\approx \omega_+\approx 2$
and $\omega_{ch}\approx 3$, and a rather strong dependence on
$N_P^{proj}$.

Fig. 1 also shows results of the HSD and UrQMD simulations
for the full 4$\pi$ acceptance for final particles, and shows the NA49-like
acceptance in the mirror rapidity interval, $-2.6<y<-1.1$ of the target
hemisphere. HSD and UrQMD both result in large values of $\omega_i$, i.e.
large fluctuations in the backward hemisphere:
in the backward rapidity interval $-2.6<y<-1.1$ (target
hemisphere) the fluctuations are much larger than those
calculated in the forward rapidity interval $1.1<y<2.6$ (projectile
hemisphere, where the NA49 measurements have been done). Even larger
fluctuations follow from the HSD and UrQMD simulations for the full
acceptance of final particles.

\section{Event-by-Event Fluctuations of Hadron Multiplicities}

The HSD and UrQMD results raise
two main questions:
\begin{itemize}
\item
What is the origin of strong fluctuations ($\omega_i$ is much larger
than 1) within the HSD and UrQMD simulations both in the full acceptance and
in the target hemisphere?
\item
Why are no large fluctuations observed in the HSD and UrQMD simulations
of the NA49 acceptance, i.e. within the projectile hemisphere?
\end{itemize}

It appears that even with the rigid centrality trigger,
$N_P^{proj}=const$, the number of nucleon participants still fluctuates
considerably.  In each sample the number of target  participants
fluctuates around its mean value, $\langle N_P^{targ} \rangle \approx
N_P^{proj}$, with the variance $V(N_P^{targ}) \equiv
\langle(N_P^{targ})^2 \rangle - \langle N_P^{targ} \rangle^2 $.
The crucial point is  that by this  event selection  one
introduces an asymmetry between projectile and target participants.
The number of projectile participants is constant by construction,
whereas the number
of target participants fluctuates.  What will be the consequences of
this asymmetry in the final observables?  As we will see later the
answer depends on dynamics or properties of the model, respectively.

 At fixed values of $N_P^{proj}$ and $N_P^{targ}$ one can introduce
 the average ($i=-,+,ch;~k=1,2,\cdots$):
 \eq{ \overline{N_i^k}~\equiv~\sum_{N_i\ge 0}N_i^k~
  P(N_i\mid N_ P^{targ},N_P^{proj}) ~,
 \label{statav}
 }
 where $P(N_i\mid N_ P^{targ},N_P^{proj})$ is the probability for producing
 $N_i$ final hadrons at fixed  $N_ P^{targ}$ and $N_P^{proj}$.
 In fact, only $N_P^{proj}$ is fixed experimentally -- hence, also
  in the HSD and UrQMD
 simulations presented in Fig.~1. The value of $N_ P^{targ}$ fluctuates,
and we denote the average over the target participants as
 \eq{
 \langle \cdots \rangle~\equiv~ \sum_{N_P^{targ}\ge 1}^{A}
 \cdots ~W(N_P^{targ}\mid N_P^{proj})~,
 \label{dynav}
 }
where $W(N_P^{targ}\mid N_P^{proj})$  is the probability for a given
value of $N_ P^{targ}$ in a sample of events with fixed number of the
projectile participants, $N_P^{proj}$. The scaled variances,
$\omega_P^{targ}$, defined as
\eq{
 \omega_P^{targ}~&\equiv~\frac{\langle\left(N_P^{targ}\right)^2\rangle~-~
 \langle N_P^{targ}\rangle^2}{
 \langle N_P^{targ}\rangle}~,
 \label{omegaPtarg}
}
give a quantitative measure of the $N_ P^{targ}$ fluctuations.

Fig.~2 presents the scaled variances $\omega_P^{targ}$ calculated
within the HSD and UrQMD models as functions of $N_P^{proj}$.
The fluctuations of $N_ P^{targ}$ are quite strong;  the
largest value of $\omega_P^{targ} = 3 - 3.5$ occurs at
$N_{P}^{proj}=20 - 30$.

\begin{figure}[t!]
\epsfig{file=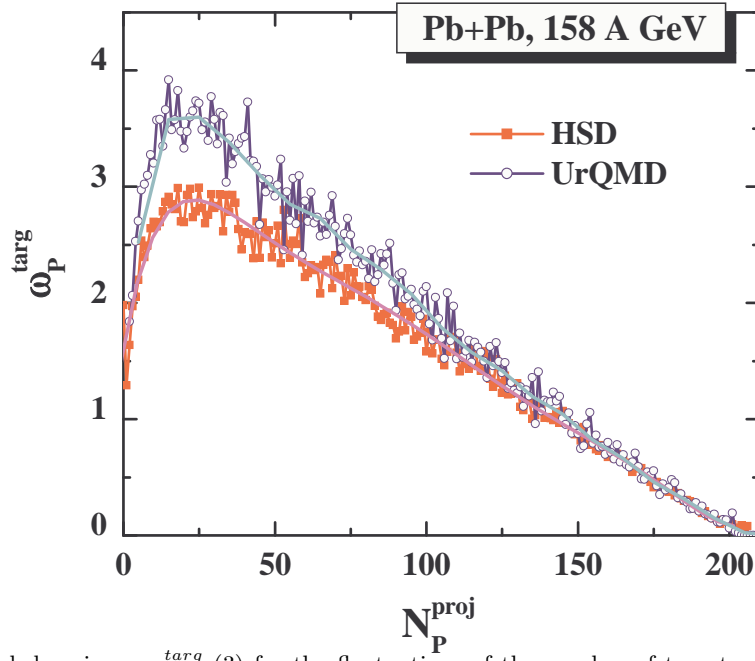,width=10cm}
\vspace{-0.5cm}
  \caption{(Color online) Scaled variance $\omega_P^{targ}$ (\ref{omegaPtarg})
  for the fluctuations of the number of target participants,
  $N_P^{targ}$.
  HSD and UrQMD simulations
  show similar behavior of $\omega_P^{targ}$
  as a function of $N_P^{proj}$.}
 \label{omegat}
 \end{figure}

 The total averaging procedure, $\langle\langle \cdots\rangle\rangle$,
performed at fixed number of projectile participants,
 $N_P^{proj}$, includes both the averaging (\ref{statav})
 and (\ref{dynav}), and can be therefore presented as
 \eq{
 \langle\langle N_i^k \rangle\rangle ~\equiv~ \langle
 \overline{N_i^k} \rangle~,
 \label{totav}
 }
 so that the total variance is:
 \eq{ Var(N_i) ~& \equiv \langle\langle N_i^2\rangle\rangle
 ~-~ \langle\langle N_i\rangle\rangle^2~ =~
 \langle \overline{N_i^2}\rangle~-~\langle \overline{N_i}\rangle^2~
 \equiv~\langle \overline{N_i^2}\rangle~-~\langle \overline{N_i}^2\rangle~
+~ \langle \overline{N_i}^2\rangle~-~\langle \overline{N_i}\rangle^2~\nonumber
\\
  &=~\langle \overline{N_i^2}~-~ \overline{N_i}^2\rangle~+~
  \langle \overline{N_i}^2\rangle~-~\langle \overline{N_i}\rangle^2~
  = ~ \omega_i^* ~\langle\overline{N_i}\rangle ~
  +~\omega_P~n_i~\langle\overline{N_i}\rangle~,
  \label{VarNi}
 }
 where
 \eq{\omega_i^* ~\equiv~ \frac{\overline{N_i^2}~-\overline{N_i}^2}
 {\overline{N_i}}~,
 ~~~~\omega_P~\equiv~\frac{\langle N_P^2\rangle~-~\langle N_P\rangle^2}
 {\langle N_P\rangle}~,~~~~ n_i~\equiv~
 \frac{\langle \overline{N_i}\rangle }{\langle N_P \rangle}~,
 \label{omegaP}
 }
and $N_P=N_P^{targ}+N_P^{proj}$, is the total number of participants.
At the last step in Eq.~(\ref{VarNi}) two assumptions have
been made. First, it is assumed that $\omega_i^*$
does not depend on $N_P$
and can be thus
taken out from the averaging, $\langle\cdots\rangle$,
in Eq.~(\ref{dynav}). The second assumption is that
the average multiplicities $\overline{N_i}$ are proportional
to the number of participating nucleons, i.e.
 $\overline{N_i}= N_P n_i$, where $n_i$ (defined in Eq.~(\ref{omegaP}))
 is the average number of particles of $i$-th type per participant.

Finally, the scaled variances, $\omega_i$ ,
can be presented as:
\eq{
\omega_i ~ \equiv ~ \frac{Var(N_i)}{\langle \overline{N_i} \rangle}~=~
\omega_i^*~+~\omega_P ~n_i~.
 \label{omegai}
 }
The total number of participants fluctuates due to the fluctuations of
$N_P^{targ}$ (the values of $N_P^{proj}$ are fixed experimentally, as well as
in the HSD and UrQMD simulations). One calculates the average values,
 $\langle N_P^{targ}\rangle\simeq N_P^{proj}$,
 and scaled variances,  $\omega_P^{targ}$,
for the target participants in both the HSD and UrQMD models (see Fig.~2).
The scaled variance $\omega_P$ (\ref{omegaP})
for the total number of participants is easily found,
$\omega_P=
\omega_P^{targ}/2$, as
only a half of the total number, $N_P$, of participants,
 i.e., $N_P^{targ}$,
does fluctuate.

\begin{figure}[h!]
 \epsfig{file=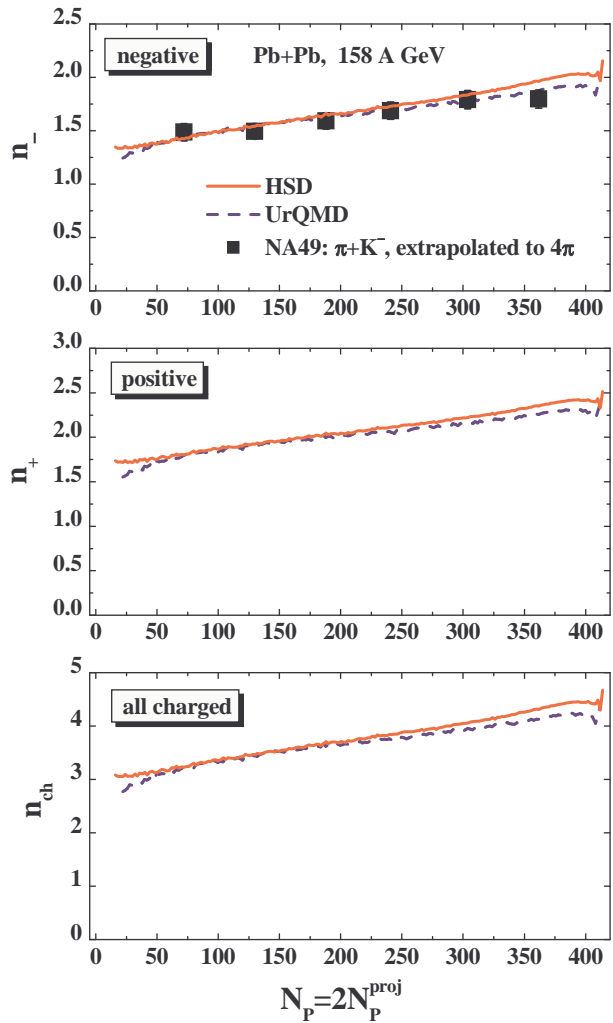,width=8cm}
  \caption{(Color online) The average particle number of $i$-th type
 ($i=$ positive, negative and all charged) per participant
  calculated within the HSD (solid lines) and UrQMD (dashed lines) models
 for full acceptance ($4\pi$).
 The squares correspond to the
 the NA49 data (extrapolated to full acceptance \cite{RNa49_QM99})
for the average $\pi+K^-$ multiplicity
 over the number of nucleon participants, using $\pi\equiv(\pi^-+\pi^+)/2$.}
 \label{ni}
 \end{figure}

Putting everything together we get:
\eq{
\omega_i ~ =~
\omega_i^*~+~\frac{1}{2}~\omega_P^{targ} ~n_i~.
 \label{omegai2}
 }
The value of $\omega_P^{targ}$ depends on $N_P^{proj}$, as shown by the
HSD and UrQMD results in Fig.~\ref{omegat}.  The average particle
number $n_i$ of $i$-th type ($i=$ positive, negative and all charged)
per participant calculated within the HSD (solid lines) and UrQMD
(dashed lines) models for full acceptance ($4\pi$) are presented in
Fig. \ref{ni}.  The squares correspond to the NA49 data (extrapolated
to full acceptance \cite{RNa49_QM99}) for the average $\pi+K^-$
multiplicity (which is an approximately 95\% of  all negatively charged
hadrons) over the number of nucleon participants, using
 $\pi\equiv(\pi^-+\pi^+)/2$.  As seen from  Fig.~\ref{ni}, both
transport models show a good agreement with each other as well as with
the extrapolated $4\pi$ NA49 data. We will use $n_i$ from Fig.
\ref{ni} for our further model calculations.

The Eq.~(\ref{omegai}) coincides with the result of the so called
'participant model' (see e.g., \cite{fluc7}), i.e. a model which treats
the A+A collision as a superposition of independent nucleon-nucleon
(N+N) interactions.  The same result (\ref{omegai}) can be obtained
within a more general framework. One assumes that a part of the initial
projectile and target energy  is converted into hadron sources.  The
numbers of projectile and target related sources are taken to be
proportional to the number of projectile and target participant
nucleons, respectively.  This results in Eq.~(\ref{omegai}).  The
physical meaning of the different sources depends on
the model under consideration
(e.g.,  wounded nucleons \cite{wnm}, strings and  resonances
\cite{HSD,UrQMD}, or the fluid cells at
chemical freeze-out, in the hydrodynamical models).
The Eq.~(\ref{omegai}) presents the final multiplicity fluctuations as a
sum of two terms: the fluctuations from one source, $\omega_i^*$, and
the contribution due to the fluctuations of the number of sources,
$\omega_P n_i$.

In peripheral A+A collisions there are only few N+N collisions, and
rescatterings are rare, so that the picture of independent N+N
collisions looks reasonable.  In this case, a hadron production source
can be associated with a N+N collision and, therefore, the fluctuations from
one source read:
\eq{
\omega_i^*~=~\omega_i^{NN}~=~\frac{\alpha_{pp}~\omega_i^{pp}~\overline{N_i}^{pp} ~+~
 \alpha_{pn}~\omega_i^{pn}~\overline{N_i}^{pn}  ~+~\alpha_{nn}
 ~\omega_i^{nn}~\overline{N_i}^{nn}}{\alpha_{pp}~\overline{N_i}^{pp} ~+~
 \alpha_{pn}~\overline{N_i}^{pn}  ~+~\alpha_{nn}
 ~\overline{N_i}^{nn}}~,
 \label{omegai*}
}
where
\eq{ \alpha_{pp}~=~Z^2/A^2=0.155~,~~~~\alpha_{pn}~=~2Z(A-Z)/A^2=0.478~,~~~~
 \alpha_{nn}~=~(A-Z)^2/A^2=0.367~
 \label{alpha}
 }
are the probabilities of proton-proton, proton-neutron, and
neutron-neutron collisions in Pb+Pb reactions (A=208, Z=82).  The
average multiplicities and scaled variances for elementary collisions
calculated within the HSD simulations at 158 GeV are equal to:
\eq{ \overline{N_{ch}}^{pp} ~& =~6.2~,~~~~
\overline{N_{ch}}^{pn}~=~5.8~,~~~~
 ~\overline{N_{ch}}^{nn}~=~5.4~,\\
\omega_{ch}^{pp} ~& =~2.1~,~~~~
 \omega_{ch}^{pn} ~=~2.4~,~~~~\omega_{ch}^{nn}~=~2.9~.
 \label{omegachNN}
}
For negatively and positively
charged hadrons, the average multiplicities
and scaled variances  in elementary reactions can be presented in terms of
corresponding quantities for all charged particles:
$\overline{N_{\pm}}=0.5(\overline{N_{ch}}\pm \gamma)$ and
$\omega_{\pm}=0.5\omega_{ch} \overline{N_{ch}}/(\overline{N_{ch}}\mp \gamma)$,
with $\gamma = 2, 1,0$ for pp, pn and nn reactions, respectively.
This yields:
\eq{\overline{N_{-}}^{pp} ~& =~2.1~,~~
 \overline{N_{-}}^{pn}~=~2.4~,~~
 ~\overline{N_{-}}^{nn}~=~2.7~,~~
\overline{N_{+}}^{pp} ~ =~4.1~,~~
 \overline{N_{+}}^{pn}~=~3.4~,~~
 ~\overline{N_{+}}^{nn}~=~2.7~,\\
\omega_{-}^{pp} ~& =~1.55~,~~
 \omega_{-}^{pn} ~=~1.5.~,~~\omega_{-}^{nn}~=~1.45~,~~
 \omega_{+}^{pp} ~ =~0.8~,~~
 \omega_{+}^{pn} ~=~1.0.~,~~\omega_{+}^{nn}~=~1.45~.
 \label{omegampNN}
 }
 From these equations one finds the HSD results for $\omega_i^*$
per N+N  collision at 158 GeV:
\eq{
\omega_{ch}^*~=~2.5~,~~~~\omega_-^*~=~1.5~,~~~~\omega_+^*~=~1.1~.~~~~
\label{omegaiNN}
}

\begin{figure}[h!]
 \epsfig{file=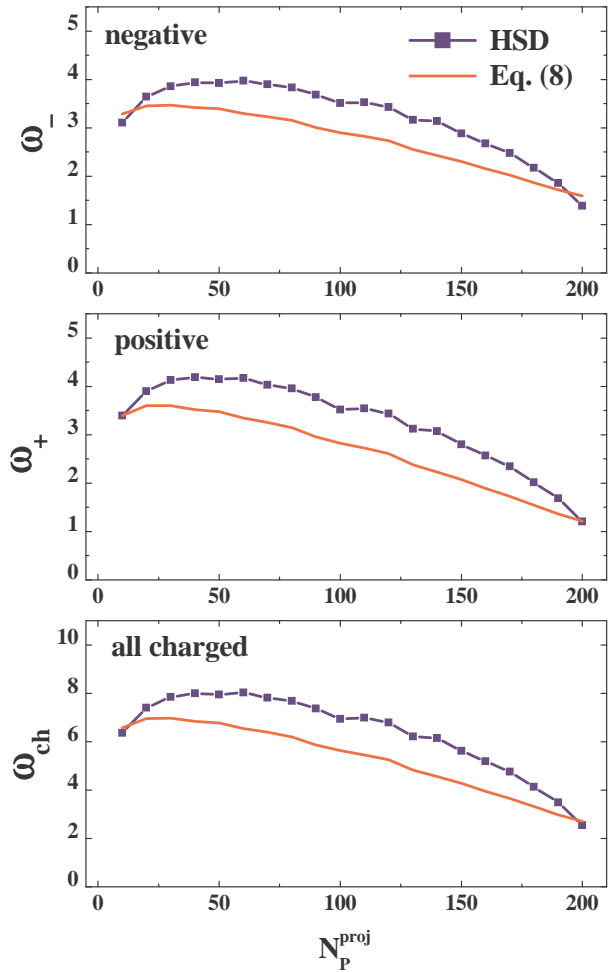,width=8cm}
  \caption{(Color online) The boxes are the results of the
  HSD simulations for $\omega_i$ in full $4\pi$ acceptance
  as functions of $N_P^{proj}$.
  The solid lines correspond to Eq.~(\ref{omegai2})
  with  $\omega_i^*$ taken from Eq.~(\ref{omegaiNN}).  }
 \label{fullacc}
 \end{figure}

The above arguments of the 'participant model' are not
applicable for central A+A collisions, where a large degree of
thermalization is expected.  In the limit of $N_P^{proj}=A$ one can
take the values of $\omega_i^*$ from the Pb+Pb data or model simulations.
In this limit, $\omega_P=\omega_P^{targ}/2\approx 0$ (see
Fig.~\ref{omegat}), and thus $\omega_i\approx\omega_i^*$. We have found
that Eq.~(\ref{omegaiNN}) gives a reasonable description of $\omega_i$
in the HSD simulations for central Pb+Pb collisions, too. Therefore, we
will use Eqs.~(\ref{omegai2}) and (\ref{omegaiNN}) for all values of
$N_P^{proj}$.
A comparison of Eq.~(\ref{omegai2}) with the HSD simulations (accepting
all final
particles) is presented in Fig.~\ref{fullacc}.

The values of $\omega_P^{targ}$
and $n_i$
are calculated within
the HSD model (see Figs.~\ref{omegat} and \ref{ni}),
and for $\omega_i^*$ we use
Eq.~(\ref{omegaiNN}). As seen from Fig.~\ref{fullacc},
there is a qualitative agreement between
Eq.~(\ref{omegai2}) and the HSD simulations. The fluctuations
of the total hadron multiplicities - generated by the HSD
dynamics - are large (the $\omega_i$ are essentially larger than 1).
The main contributions to
$\omega_i$ come from the second terms in Eq.~(\ref{omegai2}), which are
due to the
fluctuations of $N_P^{targ}$.
These fluctuations of the target nucleon participants
presented in Fig.~\ref{omegat} explain both, the large
values of $\omega_i$ and their strong dependence on $N_P^{proj}$.
Therefore, Eq.~(\ref{omegai2}) takes into account two main ingredients of
the multiplicity fluctuations in Pb+Pb collision:
a fluctuation of the particle number
created in a single N+N collision and a fluctuation in the number of nucleon
participants. Fig.~\ref{fullacc} shows that
the HSD dynamics produces even larger values of $\omega_i$
than those calculated from Eq.~(\ref{omegai2}). A very similar picture occurs
for the UrQMD model.

\begin{figure}[h!]
 \epsfig{file=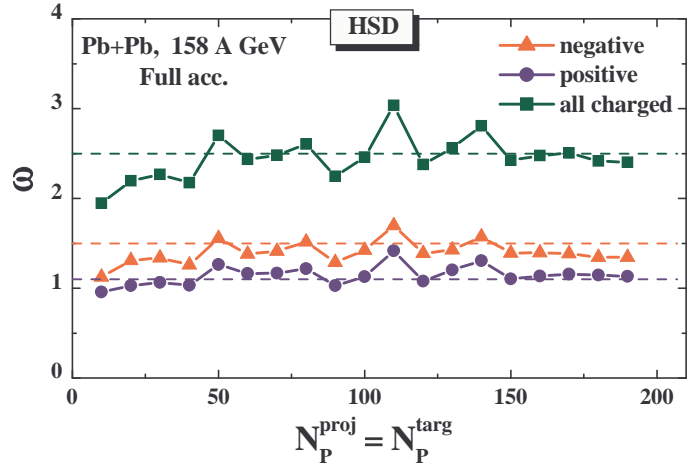,width=9cm}
  \caption{(Color online) The circles, triangles, and boxes are the results of the
  HSD simulations for $\omega_i$ in full $4\pi$ acceptance with
  $N_P^{targ}=N_P^{proj}$. This condition yields $\omega_P^{targ}=0$,
  and Eq.~(\ref{omegai2}) is reduced to $\omega_i=\omega_i^*$.
  The dashed lines correspond to
  $\omega_i^*$ taken from Eq.~(\ref{omegaiNN}).  }
 \label{w_NpNt}
 \end{figure}

Figure \ref{w_NpNt} supports the previous findings.  HSD
events
with fixed target participant number, $N_P^{targ}=N_P^{proj}$, exhibit
much smaller
multiplicity fluctuations. This is due to the fact that terms proportional to
$\omega_P^{targ}$ in
Eq.~(\ref{omegai2}) do not contribute, and $\omega_i$ become approximately
equal to $\omega_i^*$.

\section{Fluctuations in the projectile and target
hemispheres}

\begin{figure}[t!]
 \epsfig{file=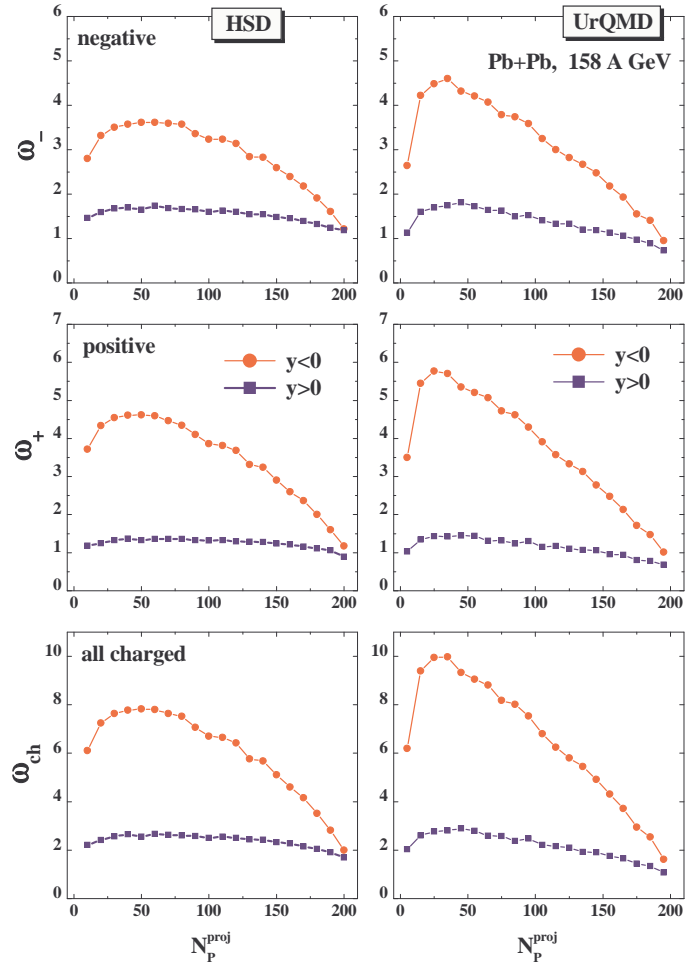,width=9cm}
  \caption{(Color online) The scaled variances
  $\omega_i$ for the projectile
  (boxes) and target (circles) hemispheres in the HSD
  (left) and UrQMD (right) simulations.}
\label{proj-targ}
\end{figure}

\begin{figure}[h!]
\epsfig{file=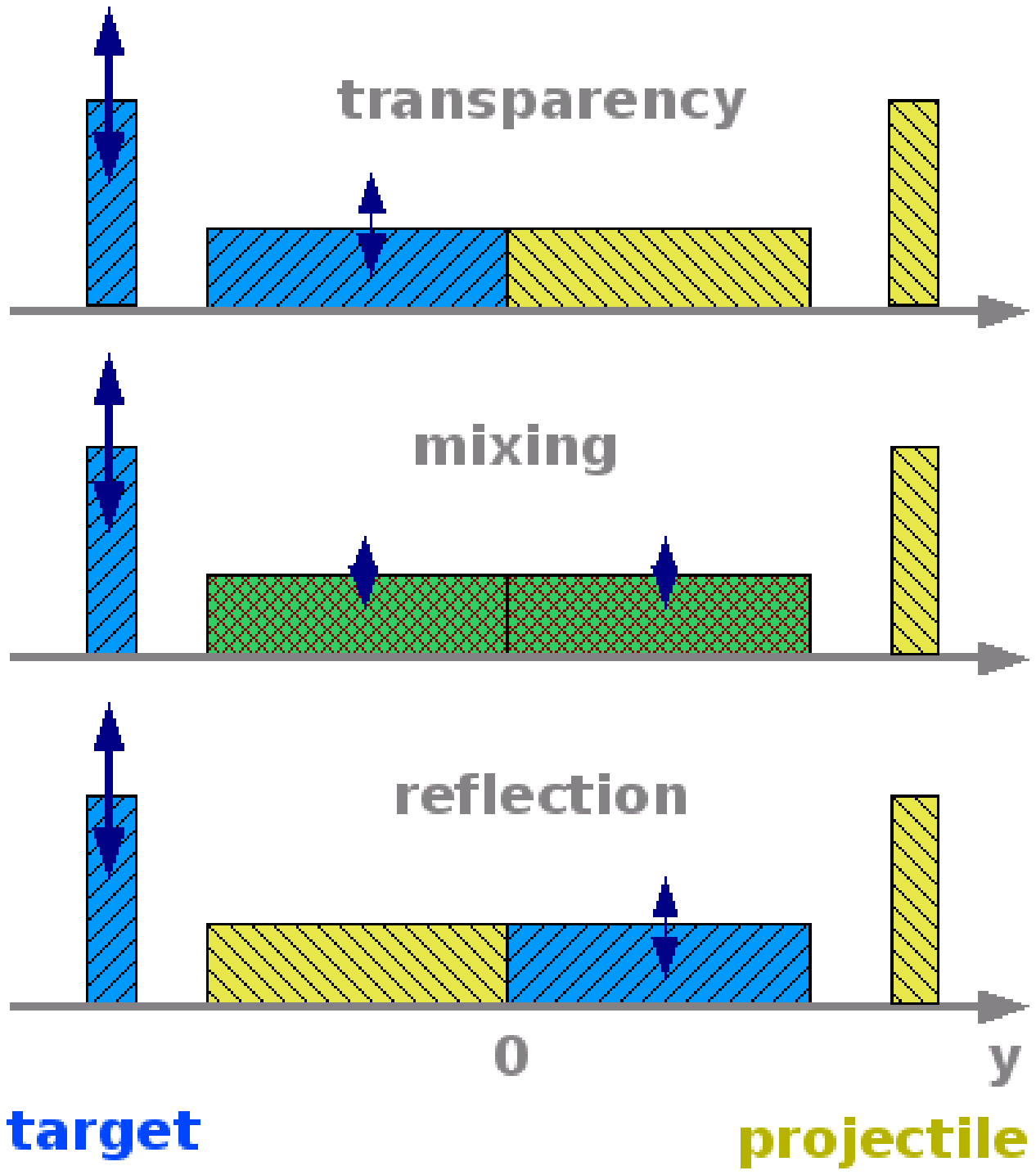,width=6.8cm}
\caption{(Color online) The rapidity distributions of
the particle production sources in  nucleus-nucleus
collisions resulting from  transparent,
mixing, and reflection models (see Ref.~\cite{MGMG} and text
for details). \label{sketch}}
\end{figure}

Let us consider now the fluctuations of the particle multiplicities in the
projectile ($y>0$) and target ($y<0$) hemispheres.  As one can see from
Fig.~2, in samples with $N_P^{proj} = const$ the number of target
participants, $N_P^{targ}$, fluctuates considerably.  Of course,
this event selection procedure
introduces an asymmetry between projectile and target participants:
$N_P^{proj}$ is constant, whereas $N_P^{targ}$ fluctuates.
Then both simulations, HSD and UrQMD,  give very different results for
the particle number fluctuations in the projectile
and target hemispheres.
 The particle number fluctuations in
the target hemispheres are much stronger (see Fig.~\ref{proj-targ})
than those in the projectile
hemispheres. There is also a strong $N_P^{proj }$-dependence of
 $\omega_i$ in the target hemisphere, which is almost absent for the
 $\omega_i$ in the projectile hemisphere. This is due to
the asymmetry between projectile and target participants.
The target participants,$N_P^{targ}$, play a quite small role for
the particle production
in the projectile hemisphere. Thus, the fluctuations of $N_P^{targ}$
have a small influence on the final multiplicity fluctuations
in the projectile hemisphere, but they contribute very strongly to
those in the target hemisphere.

Different models of hadron production in relativistic A+A collisions can be
divided into three limiting groups: transparency, mixing, and
reflection models (see Ref.~\cite{MGMG}).
The first group assumes that
the final longitudinal flows of the hadron production sources related
to projectile and target participants follow in the directions of the
projectile and target, respectively.
We call this group of models transparency (T-)models.  If the
projectile and target flows of hadron production sources
 are mixed, we call
these models the mixing (M-)models. Finally, one may even speculate
that the initial flows are reflected in the collision process. The
projectile related matter  then flows in the direction of the target
and the target related matter flows in the direction of the projectile. This
class of models we call the reflection (R-)models.  The
rapidity distributions resulting from the T-, M-, and R-models are sketched
in Fig.~\ref{sketch} taken from Ref.~\cite{MGMG}.

An asymmetry between the projectile and target participants introduced
by the experimental selection procedure can be used to distinguish
between projectile related and target related final state flows of
hadron production sources as suggested in Ref.~\cite{MGMG}.
One expects large fluctuations
of  hadron multiplicities in the domain of the target related flow and
small fluctuations in the domain of the projectile related flow.  When
both flows are mixed, intermediate fluctuations are predicted.  The
different scenarios are presented in Fig.~\ref{sketch}.
The multiplicity fluctuations measured in the
projectile momentum hemisphere clearly are larger than those measured in the
target hemisphere in T-models.  The opposite relation is predicted for
R-models, whereas for M-models the fluctuations in the projectile and
target hemispheres are expected to be the same.

  \begin{figure}[t]
 \epsfig{file=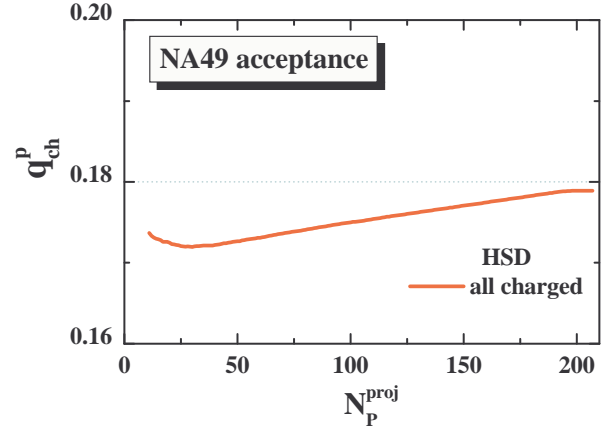,width=7.8cm}
  \caption{(Color online) The ratio of charged multiplicity within the NA49
acceptance to that in the whole projectile hemisphere. Similar results are
obtained for negative and positive hadron multiplicities. }
 \label{qi}
 \end{figure}

  \begin{figure}[!]
 \epsfig{file=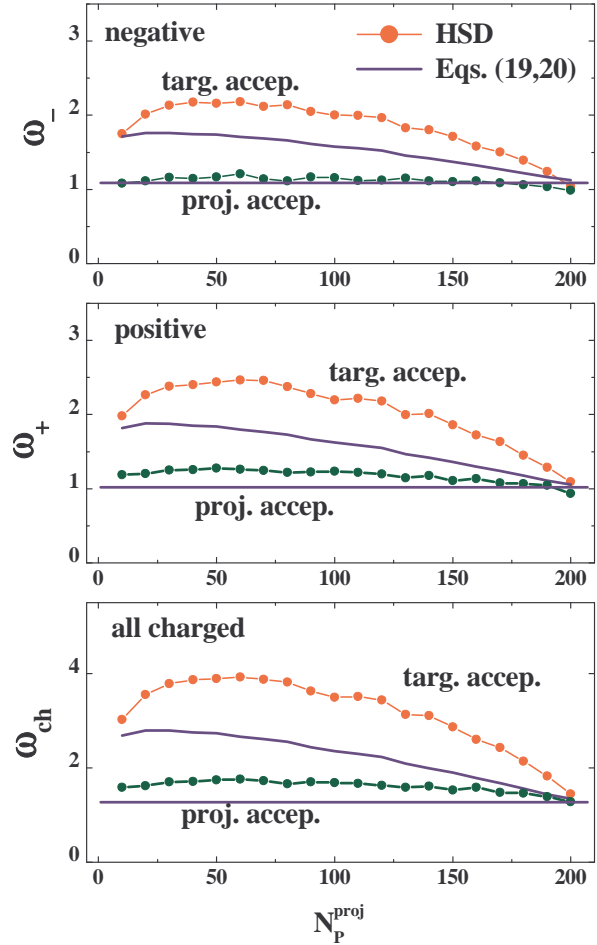,width=7.8cm}
\caption{(Color online) The HSD simulations in the NA49 acceptance in the
projectile, $1.1<y<2.6$, and target, $-2.6<y<-1.1$, hemispheres. The solid lines
correspond to Eqs.~(\ref{NA49proj},\ref{NA49targ}),
 which assume transparency of the longitudinal flows of the
hadron production sources. }
 \label{om-th-ac}
 \end{figure}

In real experiments only a fraction of all final state particles
is accepted.  In the case of weak correlations  between particles,
the scaled variances in  the limited acceptance can be calculated
( \cite{fluc7,ce1-fluc}) as $\omega_{i}^{acc}= 1 -q_i+ q_i\cdot
\omega_i$. Here the $q_i$ are the  probabilities that  particles
of type "$i$" are accepted. The $q_i$ values can be calculated as
the ratio of the average multiplicity of the $i$-th hadrons within
the given experimental acceptance inside the projectile (target)
hemisphere to the average multiplicity in the whole projectile
(target) hemisphere. The HSD values of $q_i^p\approx 0.18$ are
presented as functions of $N_P^{proj}$ in Fig.~\ref{qi} in the
NA49 acceptance (in the projectile hemisphere).

Under the above assumptions, the scaled variances of the
multiplicity distributions
in the projectile hemisphere, $\omega_i^{proj}$,
and target hemisphere, $\omega_i^{targ}$, in the T-, M- and R-models read
\cite{MGMG}:
\eq{
\omega_i^{proj}(T)~&=~1~-~q_i^p~+~q_i^p\cdot\omega_i^{*}~,~~~~
\omega_i^{targ}(T)~=~1~-~q_i^t~+~q_i^t
\cdot \left(\omega_i^{*}~+~\omega_P^{targ}~n_i
\right)~
, \label{T}\\
\omega_i^{proj}(M)~&=~\omega_i^{targ}(M)~=~1~-~q_i^{p,t}~+~q_i^{p,t}\cdot
\left(\omega_i^{*}~+ ~0.5~\omega_P^{targ}~n_i
\right)~
,\label{M}\\
 \omega_i^{proj}(R)~&=~1~-~q_i^p~+~q_i^p\cdot
\left(\omega_i^{*}~+~\omega_P^{targ}~n_i
\right)~
,~~~~ \omega_n^{targ}(R)~=~1~-~q_i^t~+~q_i^t\cdot\omega_i^{*}~.
\label{R}
}
Here $q_i^{p}$ and $q_i^{t}$ are the acceptances in the projectile and
target hemispheres, respectively.

Results presented in Fig.~\ref{proj-targ}
suggest that  HSD and UrQMD are closer to T-models. Using Eq.~(\ref{T})
the HSD simulations yield within the NA49 acceptance,
 and within the analogous acceptance in the mirror target rapidity
interval, \eq{
\omega_-^{proj}(T)~&\cong~1.09~,~~~~\omega_+^{proj}(T)~\cong~1.02~,~~~~
\omega_{ch}^{proj}(T)~\cong~1.27~,\label{NA49proj}\\
\omega_-^{targ}(T)~&\cong~1.09~+~0.18\cdot \omega_P^{targ}\cdot
n_-,~~~~ \omega_+^{targ}(T)~\cong~1.02~+~0.18\cdot
\omega_P^{targ}\cdot n_+~,
\nonumber \\
\omega_{ch}^{targ}(T)~&\cong~1.27~+~0.18\cdot \omega_P^{targ}\cdot
n_{ch}~. \label{NA49targ} }
 Here, the values of
$q_i^{p}=q_i^{t}\approx 0.18$ are taken from the HSD calculations
(Fig.~\ref{qi}), and the $\omega_i^*$ from Eq.~(\ref{omegaiNN})
are used.  The results of Eqs.~(\ref{NA49proj},\ref{NA49targ})
agree well with the HSD simulations (Fig.~\ref{om-th-ac}) for
large projectile participant number and retain the general trend
also for more peripheral collisions. Similar results are obtained
within the UrQMD simulations. Hence, both the HSD and UrQMD
approach are closer to T-models of hadron production sources.

 \begin{figure}[!]
 \epsfig{file=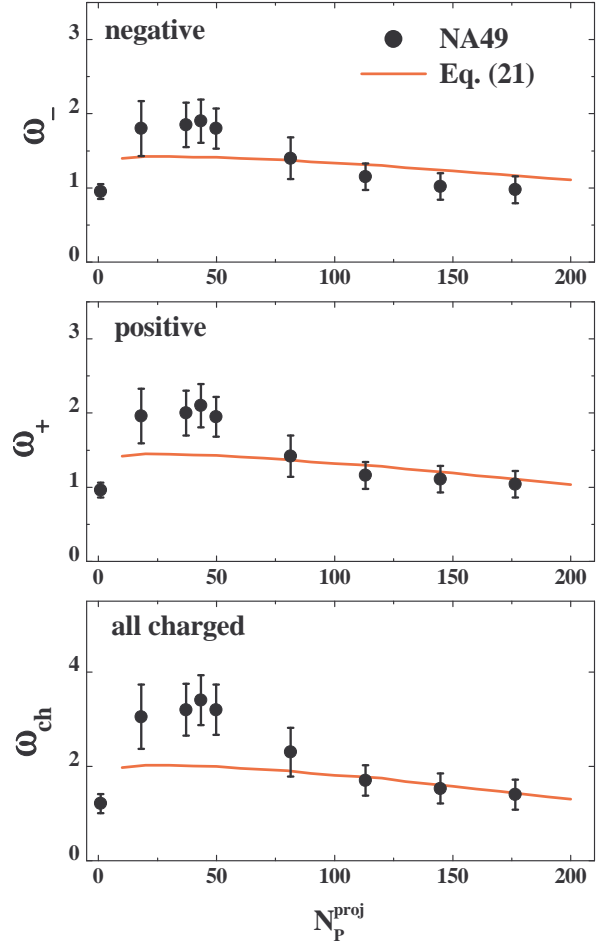,width=7.8cm}
\caption{(Color online) The solid lines correspond to
Eq.~(\ref{MNA49}) with $\omega_i^*$ (\ref{omegaiNN}),
 $\omega_P^{targ}$,
and $n_i$ taken from the HSD simulations; the points are the NA49 data.}
 \label{NA49-M}
 \end{figure}

Using Eq.~(\ref{M}) one can estimate $\omega_i$  for the NA49
acceptance in M-models. It follows: \eq{
\omega_i^{proj}(M)~=~\omega_i^{targ}(M)~= ~0.82~+~0.18\cdot
\left(\omega_i^{*}~+ ~0.5~\omega_P^{targ}~n_i
\right)~.
 \label{MNA49}
}
In Fig.~\ref{NA49-M} the results of Eq.~(\ref{MNA49}) (with
$\omega_i^*$ (\ref{omegaiNN}), $\omega_P^{targ}$, and $n_i$ taken from
the HSD simulations) are compared with the NA49 data.
Eq.~(\ref{MNA49}) for the M-model gives a  much better agreement with the
 NA49 data than Eq.~(\ref{NA49proj}) for the T-model.  The NA49 data
suggest therefore a large degree of mixing in the longitudinal flow of
the projectile- and target hadron production sources, in agreement
with suggestions formulated in Ref.~\cite{MGMG}.

A selection of collisions with a fixed number of $N_P^{proj}$ and
fluctuating number of $N_P^{targ}$ means that the projectile and
target initial flows are {\it marked in fluctuations} \cite{MGMG}
in the number of colliding nucleons. The projectile and target
related matters in the final state of collisions can be then
distinguished by an analysis of fluctuations of extensive
quantities. In the case of non-identical nuclei (different baryon
number and/or proton to neutron ratios) one can trace flows of the
conserved charges -- baryon number and electric charge -- by
looking at their inclusive final state distributions
\cite{pn1,pn2}. The analysis of the fluctuations can be applied
also to collisions of identical nuclei. Furthermore, it gives a
unique possibility to investigate the flows of particle production
sources.

\begin{figure}[!]
 \epsfig{file=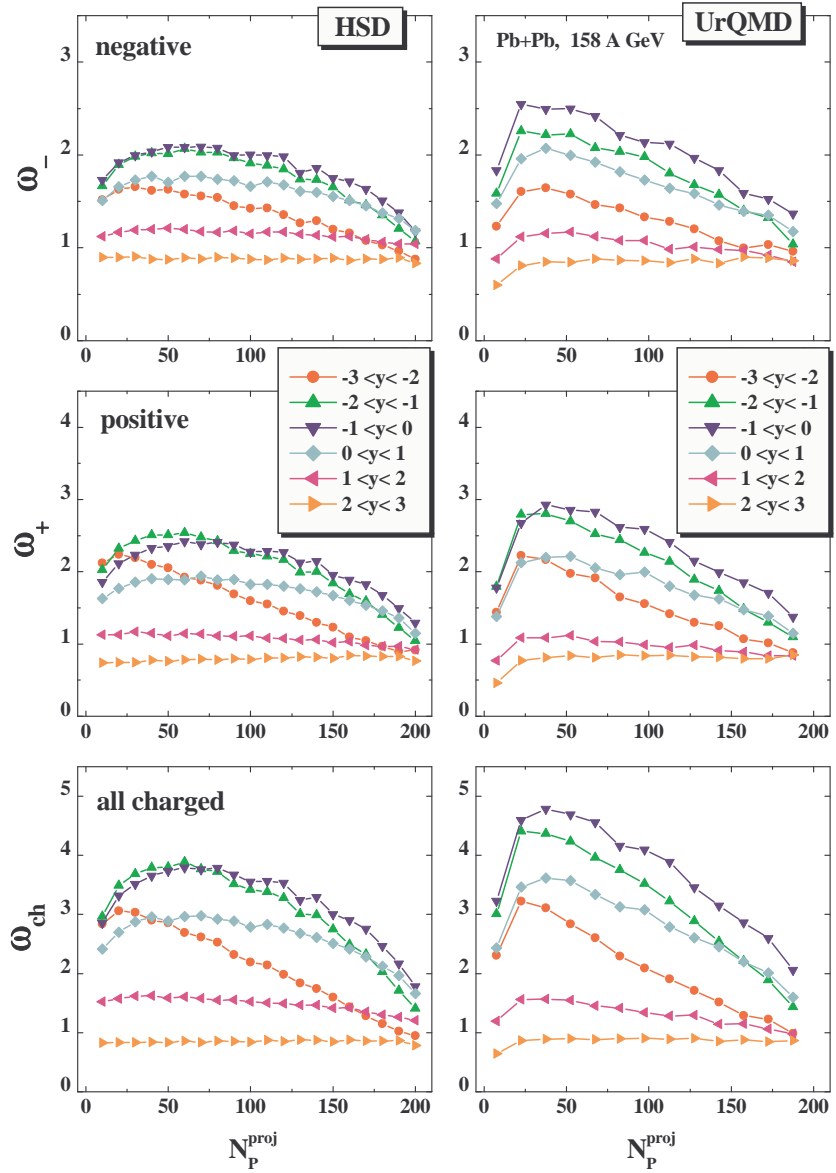,width=11cm}
\caption{(Color online) Particle number fluctuations ($\omega_-, \omega_+$, and $\omega_{ch}$)
in the HSD (left) and UrQMD simulations (right)
in different rapidity intervals in the  projectile ($y  > 0$) and target
hemispheres ($y< 0$).}
 \label{w-yint}
 \end{figure}

  \begin{figure}[h!]
 \epsfig{file=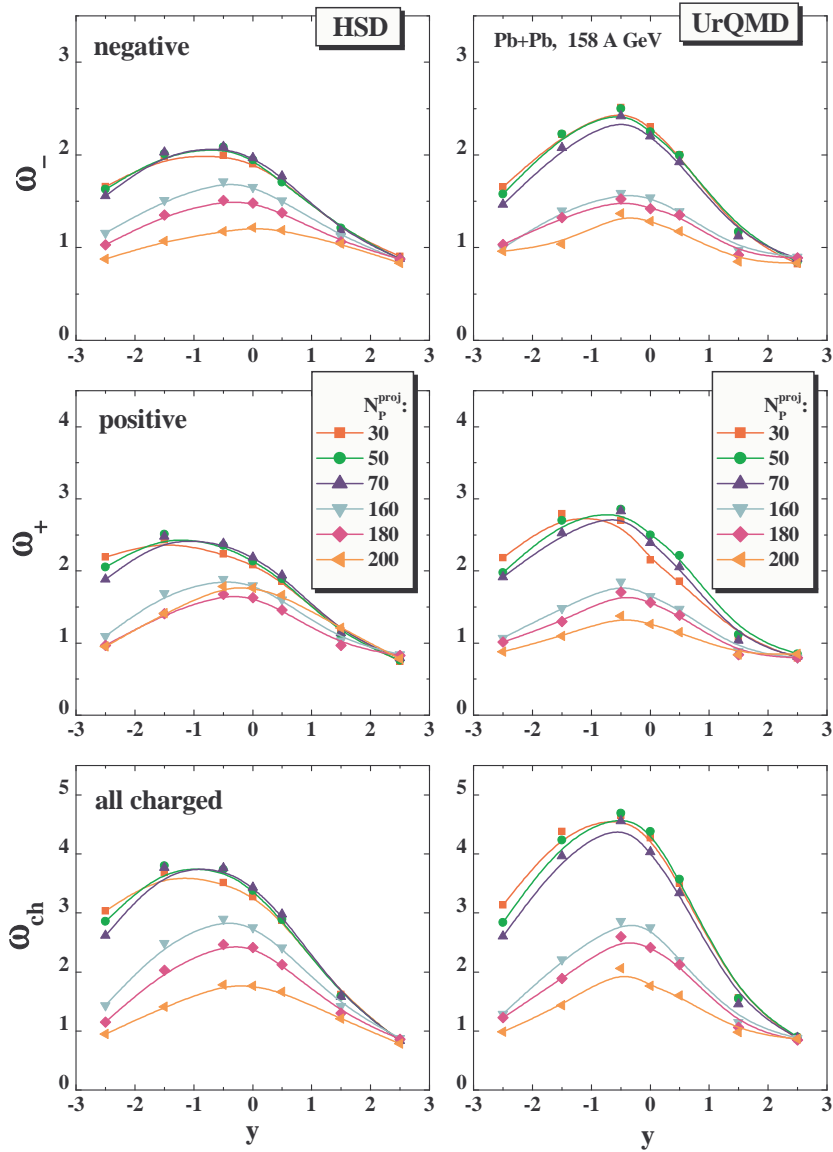,width=11cm}
\caption{(Color online) Particle number fluctuations ($\omega_-, \omega_+$, and
$\omega_{ch}$) from the HSD (left) and UrQMD (right) approaches
as a function of  rapidity $y$ for different number of projectile
participants  $N_p^{part}$.}
 \label{w-yNproj}
 \end{figure}

Fig.~\ref{w-yint} shows the particle number fluctuations ($\omega_-,
\omega_+$ and $\omega_{ch}$) in the HSD and UrQMD simulations, given in
different rapidity intervals of the projectile ($y  > 0$) and target
($y< 0$) hemispheres. The same information is presented in
Fig.~\ref{w-yNproj}, where $\omega_-, \omega_+$,
and $\omega_{ch}$ are displayed explicitly as  functions of rapidity
for different $N_p^{proj}$ values. It is clearly seen that the bias on
a fixed number of projectile participants reduces strongly the particle
fluctuations in the forward hemisphere, in particular within the NA49
acceptance ($1.1 <y <2.6$).  The fluctuations of the target participant
numbers influence strongly the hadron production sources in the target
hemispheres. They also contribute to the projectile hemisphere, but
this contribution is only important in the rapidity interval $0<y<1$,
i.e. close to midrapidity.  It turns out that this "correlation length"
in rapidity,  $\Delta y\approx 1$, as seen in Figs.~\ref{w-yint} and
\ref{w-yNproj}, is not large enough to reproduce the data.  The large
values of $\omega_i$  and their strong $N_P^{proj}$-dependence in the
NA49 data (cf. Fig. 1)  in the projectile rapidity interval,
$1.1<y<2.6$,  thus demonstrate a significantly larger amount of mixing
in peripheral reactions than generated in simple hadron/string
transport approaches.

\section{Summary and conclusions}

The event-by-event multiplicity fluctuations in Pb+Pb
collisions at 158 AGeV have been studied within the HSD and UrQMD
transport models.
The scaled variances of negative, positive, and all charged hadrons
are analyzed in minimum bias simulations for samples of events
with fixed numbers of the projectile participants, $N_P^{proj}$. This
strong centrality trigger corresponds to the trigger
of the NA49 Collaboration.

The samples with $N_P^{proj}=20-60$ show the large fluctuations of
the number of
target nucleons, $N_P^{targ}$, which participate in inelastic collisions,
$\omega_P^{targ}\ge 2$.
The final hadron multiplicity fluctuations exhibit analogous behavior,
which explains the large values of the HSD and UrQMD scaled variances
$\omega_i$ in the target hemispheres and in the full $4\pi$ acceptance.
On the other hand, the asymmetry between the projectile
and target participants -- introduced in the data samples by the trigger
condition of fixed $N_P^{targ}$ -- can be used to explore different
dynamics of nucleus-nucleus collisions by measuring the final
multiplicity fluctuations as a function of rapidity (cf. Fig. 12). This
analysis reveals that the recent NA49 data indicate a rather strong
mixing of the longitudinal flows of the projectile and target hadron
production sources. This is so
not only for central collisions -- in line with the
HSD and UrQMD approaches \cite{Weber} -- but also for rather peripheral
reactions. This sheds  new light on the nucleus-nucleus reaction
dynamics at top SPS energies for peripheral and mid-peripheral Pb+Pb
collisions. It demonstrates a significantly larger amount of mixing
than is generated in simple hadron/string transport
approaches.

The fluctuation analysis presented in this study can be performed in
the same fashion also for higher collision energies and a related
analysis in comparison to preliminary  RHIC data \cite{RHIC} will be
presented in a forthcoming study.

\begin{acknowledgments}
We would like to thank W. Cassing, M. Gazdzicki, B.~Lungwitz,
I.N.~Mishustin, St.~Mr\'owczy\'nski, M.~Rybczy\'nski,
and L.M.~Satarov for numerous discussions.
The work was supported in part by US Civilian Research and Development
Foundation (CRDF) Cooperative Grants Program, Project Agreement
UKP1-2613-KV-04.
\end{acknowledgments}


%

\end{document}